\documentclass[a4paper,conference]{IEEEtran}

\usepackage{cite}      

\usepackage{graphicx}  

\usepackage{subfigure} 

\usepackage{amsmath}    
\usepackage{mathtools}

\DeclarePairedDelimiter\floor{\lfloor}{\rfloor}

\usepackage{fancyheadings}
\begin{document}

\title{Hybrid Radiation Hydrodynamics scheme with gravity tree-based adaptive optimization algorithm}

\author{\IEEEauthorblockN{Cheryl S. C. Lau}
\IEEEauthorblockA{School of Physics and Astronomy\\
University of St Andrews\\
St Andrews, United Kingdom\\
cscl1@st-andrews.ac.uk}
\and
\IEEEauthorblockN{Maya A. Petkova}
\IEEEauthorblockA{Department of Space, Earth \\
and Environment\\
Chalmers University of Technology\\
Gothenburg, Sweden}
\and
\IEEEauthorblockN{Ian A. Bonnell}
\IEEEauthorblockA{School of Physics and Astronomy\\
University of St Andrews\\
St Andrews, United Kingdom}}


\maketitle

\begin{abstract}
Modelling the interaction between ionizing photons emitted from massive stars and their environment is essential to further our understanding of galactic ecosystems. We present a hybrid Radiation-Hydrodynamics (RHD) scheme that couples an SPH code to a grid-based Monte Carlo Radiative Transfer code. The coupling is achieved by using the particle positions as generating sites for a Voronoi grid, and applying a precise mapping of particle-interpolated densities onto the grid cells that ensures mass conservation. The mapping, however, can be computationally infeasible for large numbers of particles. We introduce our tree-based algorithm for optimizing coupled RHD codes. Astrophysical SPH codes typically utilize tree-building procedures to 
sort particles into hierarchical groups (referred to as \textit{nodes}) for evaluating self-gravity. Our algorithm adaptively walks the gravity tree and transforms the extracted nodes into pseudo-SPH particles, which we use for the grid construction and mapping. This method allows for the temporary reduction of fluid resolution in regions that are less affected by the radiation. A neighbour-finding scheme is implemented to aid our smoothing length solver for nodes. We show that the use of pseudo-particles produces equally accurate results that agree with benchmarks, and achieves a speed-up that scales with the reduction in the final number of particle-cell pairs being mapped.
\end{abstract}

\section{Introduction} \label{intro}

The SPH method remains widely used in astrophysics to simulate the dynamical evolution of Giant Molecular Clouds. We model the cloud gas as compressible flows and study the structures that form under mutual gravitational attractions, within which the stars are born. However, as the more massive stars are formed, they emit significant amount of ionizing UV radiation which modifies the local environment. These high-energy photons strip hydrogen and helium atoms of electrons and further heat them to approximately \(10^4\) K, forming regions of hot ionized gas that expand due to over-pressure, known as \textit{HII regions} (see e.g.\cite{osterbrock74}). Incorporating radiation models into hydrodynamical simulations is therefore crucial to investigate their effect on the on-going star formation. 

One of the most commonly implemented radiation algorithms in astrophysical SPH codes is \textit{ray-tracing} (e.g. \cite{kesseldeynetburkert00}\cite{dale07c}). This method works by drawing lines between the source and the surroundings, along which we solve the radiative transfer equations. Whilst being intuitive, it neglects the scattering, absorption and re-emission processes that photons can undergo. The problem is further complicated by the stochasticity in such processes. To the contrary, the Monte Carlo Radiative Transfer (MCRT) technique overcomes these limitations by explicitly simulating these random events. 

MCRT (e.g. \cite{wood04}\cite{harries11}) is a grid-based method that models radiation by discretizing the source into photon packets and propagating them through a density field. Cell densities determine the optical depths. This method follows each packet as it travels, changes direction or disappears, depending on the physical likelihood of each process. Each cell accumulates the path lengths of photons passing through and, by the end, estimates the ionic fraction within its domain. The release of packets is iterated until the ionization structure converges. Despite its computational expense, MCRT holds great advantage in capturing the micro-physics of radiation effects. 

This proceeding is organized as follows: Section \ref{coupling} describes our method to couple MCRT to SPH as first developed by \cite{petkova21}. We outline the \textit{Exact} mapping method \cite{petkova18} for transferring fluid properties between particle- and grid-based models. Section \ref{pseudo_parts} presents our algorithm for setting up the \textit{pseudo-particles}. These particles serve as the alternative sampling points of the underlying fluid, allowing resolutions to be varied during the MCRT without affecting the hydrodynamics. We present the benchmark results in Section \ref{benchmark} and justify the benefit of adopting pseudo-particles with runtime tests. Finally, in Section \ref{applications}, we apply this RHD scheme to a simulation of a star-forming molecular cloud.

\section{SPH-MCRT Coupling} \label{coupling}

\subsection{The Radiation-Hydrodynamics (RHD) scheme} \label{rhd_scheme}

The SPH code {\scshape Phantom}\cite{phantom18}\footnote{{\scshape Phantom} is dedicated to astrophysical compressible fluids. The code employs equal mass particles with variable smoothing lengths.} is used in our scheme. At each hydro step, we pass the particles' positions, masses and smoothing lengths over to the grid-based MCRT code {\scshape CMacIonize} \cite{vandenbrouckewood18}. The massive stars formed within the SPH simulation are set as the sources of ionizing radiation. We first initialize a Voronoi grid with generating sites coinciding with the input particles' locations\footnote{Grid generating sites do not necessarily overlap with particle positions since the Exact mapping method allows an arbitrary number of particles in each cell. We merge the tightly-packed cells to prevent a numerical error.}. We apply 5 Lloyd iterations \cite{lloyd82} to correct the elongated cells. Fluid densities, as interpolated from the SPH particles, are subsequently mapped onto the grid to obtain the density of each Voronoi cell (see Section \ref{mapping}). The MCRT simulation is then carried out on this density grid to compute the cells' ionic fractions. 

Reversing the procedures described above, we now map the computed ionization structure back onto the SPH particles. The ionic fraction of each gas particle provides the number densities of free electrons and protons, with which we determine the heating and cooling rates. The temperature of the particle is then evolved alongside other thermal processes within the SPH simulation. Note that only the ionized particles are heated accordingly; neutral particles are unaffected by this coupling.

\subsection{Voronoi mapping} \label{mapping}

\begin{figure}
    \includegraphics[width=3.0in]{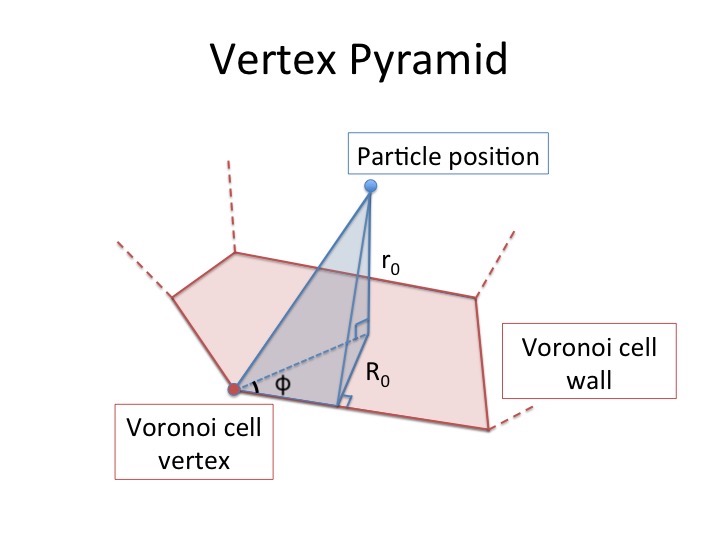}
    \caption{Illustration of a vertex pyramid used to break up the volume of a Voronoi grid cell (from \cite{petkova18}).}
    \label{fig:pyramids}
\end{figure}

The mean density of cell `$i$' is given as the mass contained within the cell divided by the cell volume ($V_i$). The cell mass can be expressed as the integral of the density over the cell volume, and we can link this to the SPH formalism in the following way \cite{petkova21}:
\begin{eqnarray}
\rho_{i} & =  & \frac{1}{V_i} \int_{V_i}  \rho(\mathbf{r'}) \mathrm{d}V' \\
 & = & \frac{1}{V_i} \int_{V_i}  \sum_{a=1}^N m_a W(|\mathbf{r'}-\mathbf{r}_a|,h_a) \mathrm{d}V'\\
 & = & \frac{1}{V_i} \sum_{a=1}^N m_a \int_{V_i}  W(|\mathbf{r'}-\mathbf{r}_a|,h_a) \mathrm{d}V'.
\label{eq:exact}
\end{eqnarray}
In the above, $W$ is the kernel function, and $m_a$, $\mathbf{r}_a$ and $h_a$ are the mass, position and smoothing length of particle `$a$' respectively. Equation (\ref{eq:exact}) demonstrates that the cell density depends on the integral of the kernel function over the volume of the cell, which is typically an irregularly-shaped polyhedron. 

To make the method applicable to a range of cell shapes, we divide the cell volume into a collection of pyramids, which we refer to as vertex pyramids (see Fig.~\ref{fig:pyramids}). A vertex pyramid is characterised by the orthogonal distance from the particle position to the plane of the relevant cell wall ($r_0$), the orthogonal distance from the projection point of the particle position on the wall to the relevant cell edge ($R_0$), and the angle between the edge and the line connecting the vertex and the projection point ($\phi$). Since a cell vertex connects multiple walls and edges, it participates in multiple vertex pyramids. The integral of $W$ over the volume of a vertex pyramid can be expressed analytically \cite{petkova18}, and it depends only on $r_0$, $R_0$ and $\phi$, and the functional form of $W$ (see Section~4.6 and Appendix~A of \cite{petkova18b}). Due to the complexity of the analytic expressions, we pre-compute the integral and interpolate between the values. This ensures mass conservation between the SPH and Voronoi grid representations within $0.6\%$ \cite{petkova18b}.

After performing the MCRT step, each cell has an assigned ionic fraction $f_{\mathrm{ion},i}$. We obtain the ionic fraction of a particle `$a$' using \cite{petkova21}
\begin{equation}
f_{\mathrm{ion},a} = \sum_{i=1}^{N} f_{\mathrm{ion},i} \int_{V_i}  W(|\mathbf{r'}-\mathbf{r}_a|,h_a) \mathrm{d}V',
\label{eq:exact-inv}
\end{equation}
which ensures that the ionized mass is conserved.

\section{Pseudo-Particles} \label{pseudo_parts}

From (\ref{eq:exact}), carrying out density-mapping requires the position, mass and smoothing length of each fluid interpolation point. We compute these quantities for the pseudo-particles with methods described in the following sections. 

\subsection{Gravity tree} \label{tree}

Self-gravity in SPH is typically solved on a tree. As gravity scales inversely with the square of separation, direct summation of forces becomes unnecessary for distant particles whose contributions are less significant. Hence, far-field gravity can be evaluated by treating such particles collectively as a group, with group size determined by its distance to the particle in concern, reducing the complexity from \(\mathcal{O}(N^2)\) to \(\mathcal{O}(N\log{N})\). 

Tree-build refers to the process of grouping particles in a hierarchical manner for this purpose. Various types of gravity trees have been developed over the past decades in efforts to optimize performance (e.g. \cite{barneshut86}\cite{press86}\cite{benz90}). In this proceeding, we specifically refer to the \textit{k}d-tree implemented in {\scshape Phantom}, developed based on the Recursive Coordinate Bisection (RCB) tree proposed by \cite{gaftonrosswog11}. 

\begin{figure}
    \centering
    \includegraphics[width=2.6in]{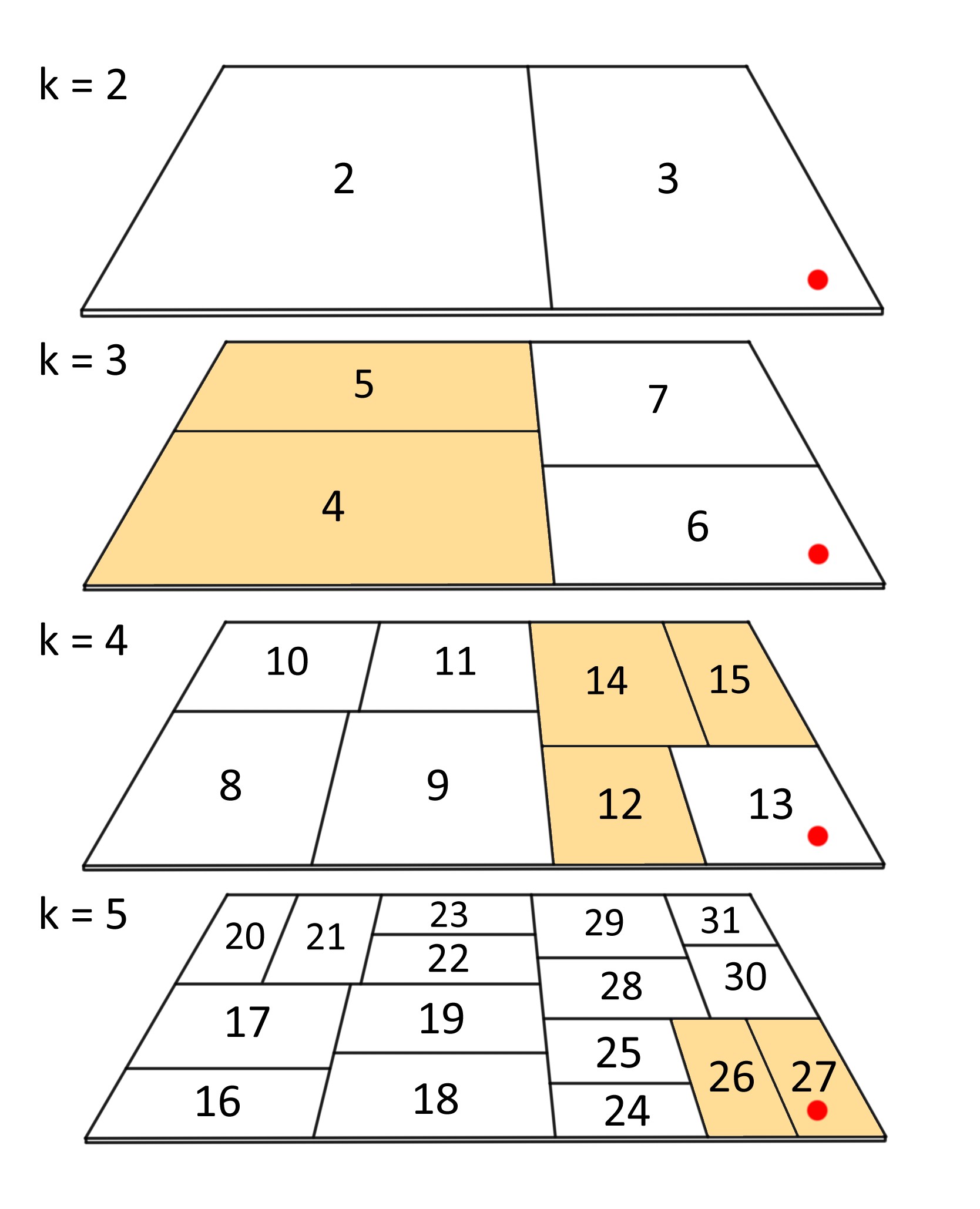}
    \caption{Illustration of the recursive splitting of simulation domain to build a gravity \textit{k}d-tree. The 3-D space is represented as a 2-D plane, and each plane in the stack represents a tree level. The sub-domains (nodes) are labelled according to the convention demonstrated in Fig. \ref{fig:tree_structure}. For illustration purpose, tree-build is terminated at the \(5^\mathrm{th}\) level. Consider a target particle at the position indicated by the red dot. Only the coloured nodes are considered in its gravitational force evaluation. }
    \label{fig:tree_layers}
\end{figure}

\begin{figure}
    \centering
    \includegraphics[width=3.3in]{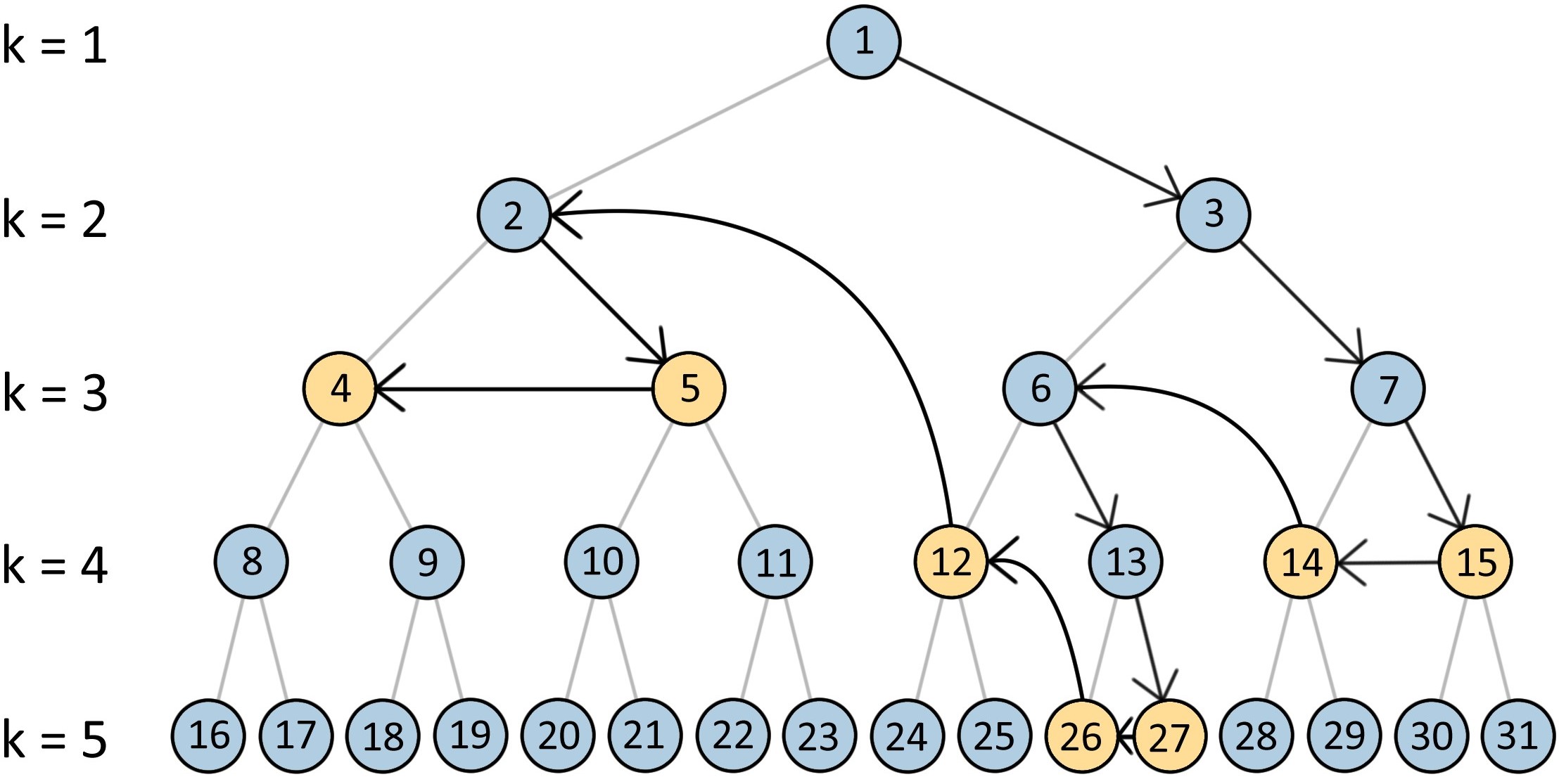}
    \caption{Structure of \textit{k}d-tree that corresponds to the example shown in Fig. \ref{fig:tree_layers}. The labelling convention enables node relations to be easily recovered. Arrows indicate an example tree-walk sequence with opening criteria defined with respect to the target particle (red dot). Nodes in orange are those which would have been \textit{accepted} during this traversal. Their physical domains are also coloured in orange in Fig. \ref{fig:tree_layers}.  }
    \label{fig:tree_structure}
\end{figure}

The RCB tree is a top-down binary tree built by recursively splitting the simulation domain into two. The split is through the centre of mass of the constituting particles to maintain load balancing, and along the longest axis of the cell to avoid elongation. We illustrate this in Fig. \ref{fig:tree_layers} and Fig. \ref{fig:tree_structure}. The sub-divided domains are referred to as \textit{nodes}; the entirety of the simulated space is the \textit{root} and the bottom-most nodes are the \textit{leaves}. Position of a node is defined to be its centre of mass, and size \(s_\mathrm{node}\) is defined to be the radius to its furthest particle. In {\scshape Phantom}, by default, the splitting procedure ends when the leave contains less than 10 particles. 

The nodes are labelled such that their indices carry information of local proximity (see Fig. \ref{fig:tree_structure}). For example, a node with index \(n_a\) is located on level \(k_a \equiv \floor*{\log_{2}{n_a}}\); its parent \(n_b\) on some higher level \(k_b\) can be determined with the formula \(n_b = \floor*{n_a / 2^{k_a - k_b}}\). This labelling system gives rise to a set of convenient arithmetic rules that allow node relations to be recovered without extra memory consumption. We utilize these relations in Section \ref{neighfind} for our neighbour-find algorithm.

To evaluate the gravitational forces acting on a certain particle, we traverse the tree. This procedure is termed \textit{tree-walk} and we hereafter specifically refer to the depth-first search algorithm implemented in {\scshape Phantom}. The arrows in Fig. \ref{fig:tree_structure} illustrates an example traversal. Tree-walk begins from the root. We go through a series of \textit{tree opening criteria} to decide whether or not this node needs to be resolved into its constituents to ensure accuracy. If yes, we examine its right child. This downward search is continued until the opening criterion is no longer met, we \textit{accept} the node as it is, and compute its force contribution via a multipole expansion up to quadrupole order (e.g. \cite{benz90}). Thereupon, we turn left and look for the adjacent branch that is yet to be traversed. The tree-walk is terminated when all branches have been visited. 

Note in particular that the domains of the accepted nodes extracted from tree-walks are effectively a re-tessellation of the simulation space, implying that total mass must be conserved. Like SPH particles, nodes also represent fluid parcels, only of different sizes and masses. Hence, converting these nodes into pseudo-SPH particles does not modify the underlying fluid, but only its resolution. Replacing the non-ionized SPH particles with pseudo-particles within the SPH-MCRT interface serves as an elegant way to optimize the coupling.

\subsection{Adaptive tree-walk} \label{tree_walk}

\begin{figure*}
    \centering
    \includegraphics[width=5.15in]{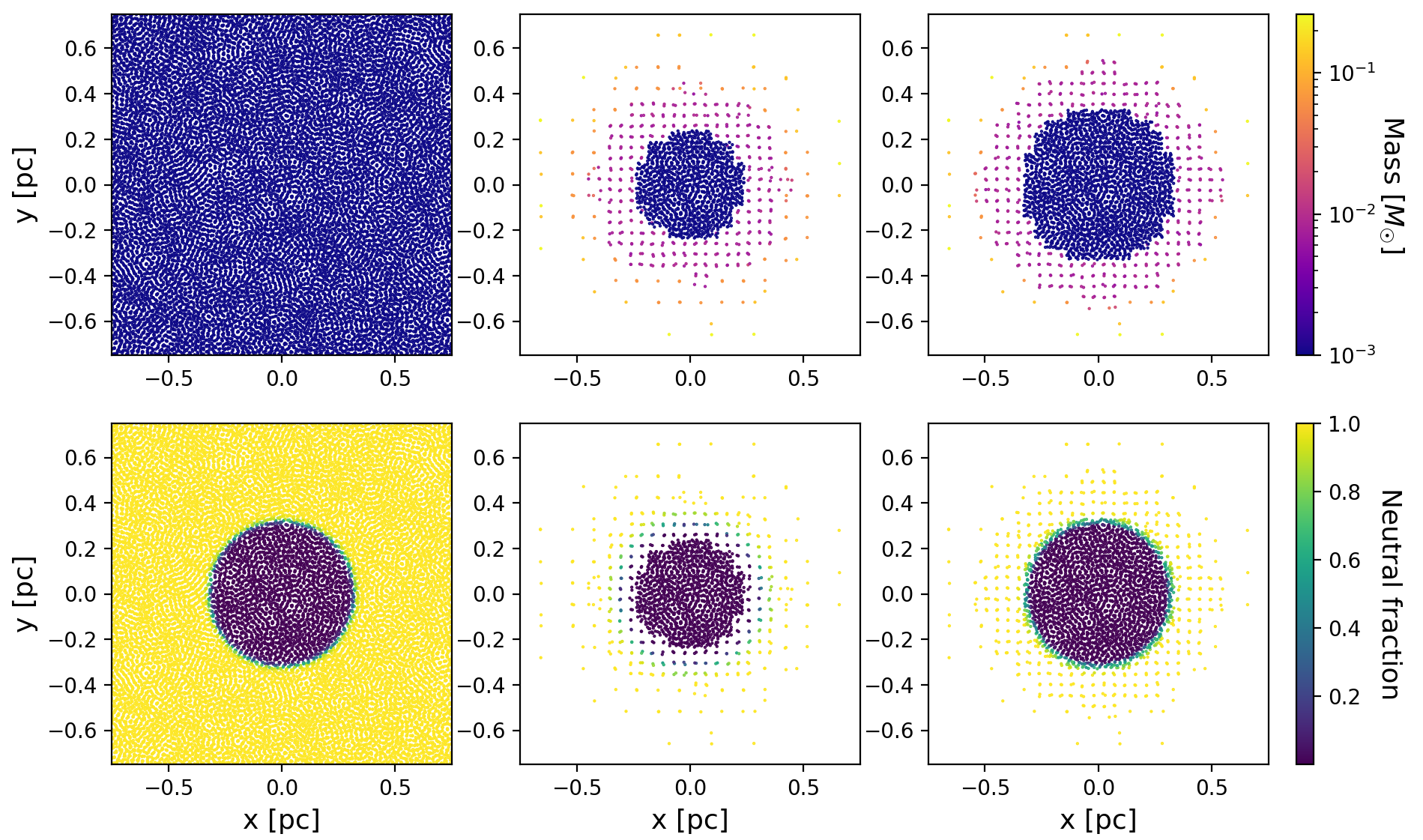}
    \caption{Slices from a 3-D simulation of an ionizing stellar source placed in a uniform medium (see Section \ref{benchmark} for details on setup). Left panels illustrate the case with all SPH particles. Middle panels illustrate the pseudo-particles obtained from a tree-walk, with \(r_{\mathrm{part}} = 0.2 \ \mathrm{pc}\) and \(r_{\mathrm{leaf}} = 0.4 \ \mathrm{pc}\). Right panels show the same but with the ionization front just resolved after the adaptive tree-walk iterations. Color scales in the upper panels indicate particle mass in units of \(M_{\odot}\) (\(1.989 \times 10^{30} \ \mathrm{kg} \)) and bottom panels indicate their neutral fractions. The sharp transition zone between ionized and neutral domains defines the ionization front. Length scales in units of \(\mathrm{pc}\) \((3.086 \times 10^{16} \ \mathrm{m})\). }
    \label{fig:node_illus}
\end{figure*}

The problem comes down to \textit{how} to walk the tree such that all domains are at their optimal resolutions. First, to ensure that the HII region is fully resolved, we define a threshold radius \(r_{\mathrm{part}}\) around the ionizing sources within which we must open the leaves and extract the individual SPH particles\footnote{We hereafter consider these SPH particles as constituents of the full set of pseudo-particles that we pass to the grid code. Note also that this treatment is only necessary if the leaves contain more than one particle.}. For the outer non-ionized regions, we follow the node opening criterion used for gravity computation, that is, to define an opening angle \(\theta\), given by 
\begin{eqnarray} \label{eq:opening_angle}
    \theta^2 <  \left( \frac{s_\mathrm{node}}{r_\mathrm{node}} \right)^2,
\end{eqnarray}
where \(s_\mathrm{node}\) is the size of the node and \(r_\mathrm{node}\) is its distance to the ionizing source. The angle \(\theta\) is a parameter set between 0 and 1. If the angle subtended by the node to the source is greater than \(\theta\), we open the node. Hence, the smaller the \(\theta\), the more pseudo-particles would be extracted. We also define another threshold radius \(r_{\mathrm{leaf}}\) such that, in the annulus between \(r_{\mathrm{leaf}}\) and \(r_{\mathrm{part}}\), leaves must be extracted to maintain some continuity in the distributions. An example is illustrated in Fig. \ref{fig:node_illus}. In situations where multiple sources are present, we walk the tree only once but allow the opening criteria to take all sources into account. 

The next objective is to adapt the pseudo-particle resolutions to expand with the ionized region. This is achieved through iterative calls to the MCRT simulation \textit{within} a hydro step. We perform checks after retrieving results from the MCRT code to see whether or not the ionized pseudo-particles are sufficiently small in size. This criterion is governed by the function 
\begin{eqnarray} \label{eq:resol_param}
    f_{\mathrm{neu,limit}} = \frac{1}{K} \left( -\frac{1}{s_\mathrm{node} / s_\mathrm{root}} + K \right),
\end{eqnarray}
where \(f_{\mathrm{neu,limit}}\) is the neutral fraction threshold, \(s_\mathrm{node}\) is the node size, \(s_\mathrm{root}\) is the size of the root node, and \(K\) is a free parameter for controlling the resolution of the partially ionized regions. Nodes with neutral fraction beneath its \(f_{\mathrm{neu,limit}}\) are considered under-resolved. 

If one or more nodes did not pass the checking criterion in (\ref{eq:resol_param}), we store their labels and open them in the next iteration. If even the leaves fail, we simply increase \(r_{\mathrm{part}}\) (with \(r_{\mathrm{leaf}}\)) as we do not perform resolution checks on the SPH particles. The MCRT simulation is then re-run with a new set of pseudo-particles until all nodes pass. Only once the code is satisfied with the current resolutions, we proceed to update the particles' internal energies. Performing such trial and error process within a step eliminates the need to predict the growth of the ionization structure, nor delaying the `response' to later timesteps which hinders the modelling accuracy. 

\begin{figure}
    \centering
    \includegraphics[width=3.05in]{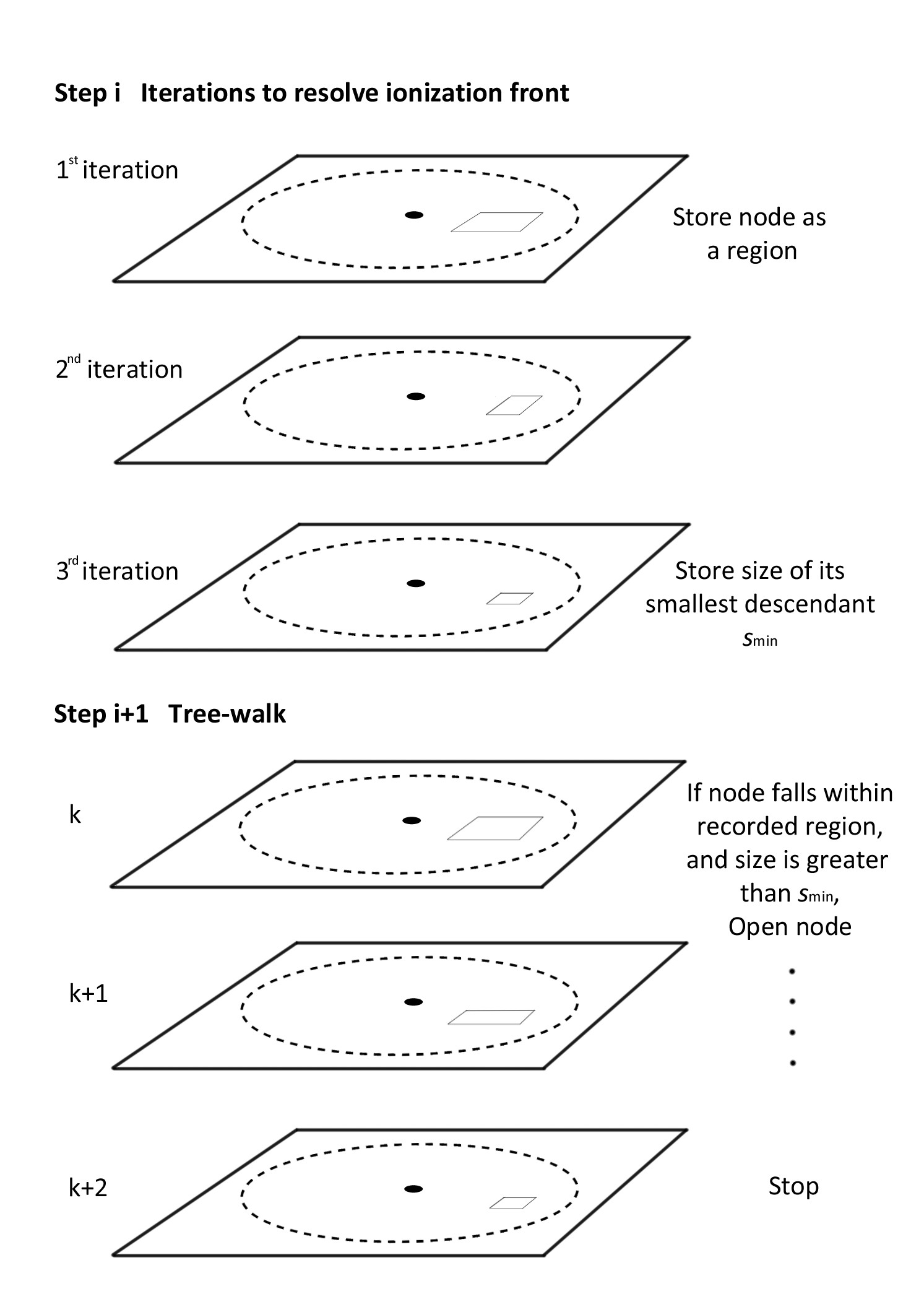}
    \caption{Illustration of the method to restore the distribution of pseudo-particles after the tree rebuilds. The 3-D simulation space is represented as 2-D planes. The circle in dotted lines is the HII region and the black dot is the ionizing source. Grey squares represent arbitrary nodes within the ionized region. }
    \label{fig:store_for_nextstep}
\end{figure}

To avoid repeating the iterations in subsequent steps, an algorithm was designed to let the code `remember' how high the pseudo-particles were on the tree. Keeping in mind that trees could rebuild as particles move, we ought to use the nodes' physical properties rather than their labels. The method is depicted in Fig. \ref{fig:store_for_nextstep}. Consider a node that failed the checks in the initial trial. This node effectively draws a spatial region where higher resolutions are required, and we record it. As the node iteratively resolves into its constituents, we, in the meantime, store the minimum size of its descendants. In the next timestep, if a node falls within the boundaries of this recorded region, we open it and advance down the tree until the node size becomes comparable to that of its previously-recorded smallest descendant. 

This algorithm does not guarantee reproducing the previous set of pseudo-particles, but it is capable of restoring the overall distribution and immediately suppressing the iterations. In the rare occasion that an ionized node becomes shielded from the stellar source, we revert the procedure by simply removing the node from the list of stored regions, pushing the pseudo-particles back up on the tree.

\subsection{Smoothing lengths} \label{h_solver}

In analogous way to the SPH formulation, we set the smoothing lengths of the pseudo-particles to correlate with their number densities \(n_{\mathrm{node}}\), giving
\begin{eqnarray} \label{eq:node_h}
    h_\mathrm{node} = h_\mathrm{fact,node} \ n_{\mathrm{node}}^{-1/3},
\end{eqnarray}
where
\begin{eqnarray} \label{eq:node_n}
    n_{\mathrm{node}} = \sum_{b_{\mathrm{node}}} W(|\textbf{r}_\mathrm{node} - \textbf{r}_{b,\mathrm{node}}|,h_{\mathrm{node}}).
\end{eqnarray}
Substituting (\ref{eq:node_n}) into (\ref{eq:node_h}), \(h_{\mathrm{node}}\) can be solved using simple root-finding algorithms. We adopt a \(M_4\) cubic spline kernel, and we set \(h_\mathrm{fact,node} = 1.1\) for each pseudo-particle to have around 55 neighbours. We let the SPH particles extracted near the source(s) (within \(r_{\mathrm{part}}\)) retain their original smoothing lengths, and thus they are not being passed into this smoothing length solver. However, considering that voided regions can lead to erroneously large smoothing lengths for the bordering pseudo-particles, we temporarily substitute these SPH particles with their leaves. 

Since pseudo-particles originate from distinct tree levels, their contrast in distributions indicates the necessity to implement multiple back-up root-finding methods. As a first attempt, the code solves for \(h_\mathrm{node}\) with the Newton-Raphson method. If the solution fails to converge, we apply the bisection method. If both attempts fail, we estimate its resolution length with 
\begin{eqnarray} \label{eq:h_estimate}
    h_\mathrm{node} = h_\mathrm{fact,node} \left( 2 s_\mathrm{node} \right).
\end{eqnarray}
Equation (\ref{eq:h_estimate}) arises from the assumption that the number density of a node is roughly equal to the inverse of the volume of its domain. We also use (\ref{eq:h_estimate}) to provide the initial guesses for the root-finding algorithms.

\subsection{Neighbour-find} \label{neighfind}

\begin{figure}
    \centering
    \includegraphics[width=3.35in]{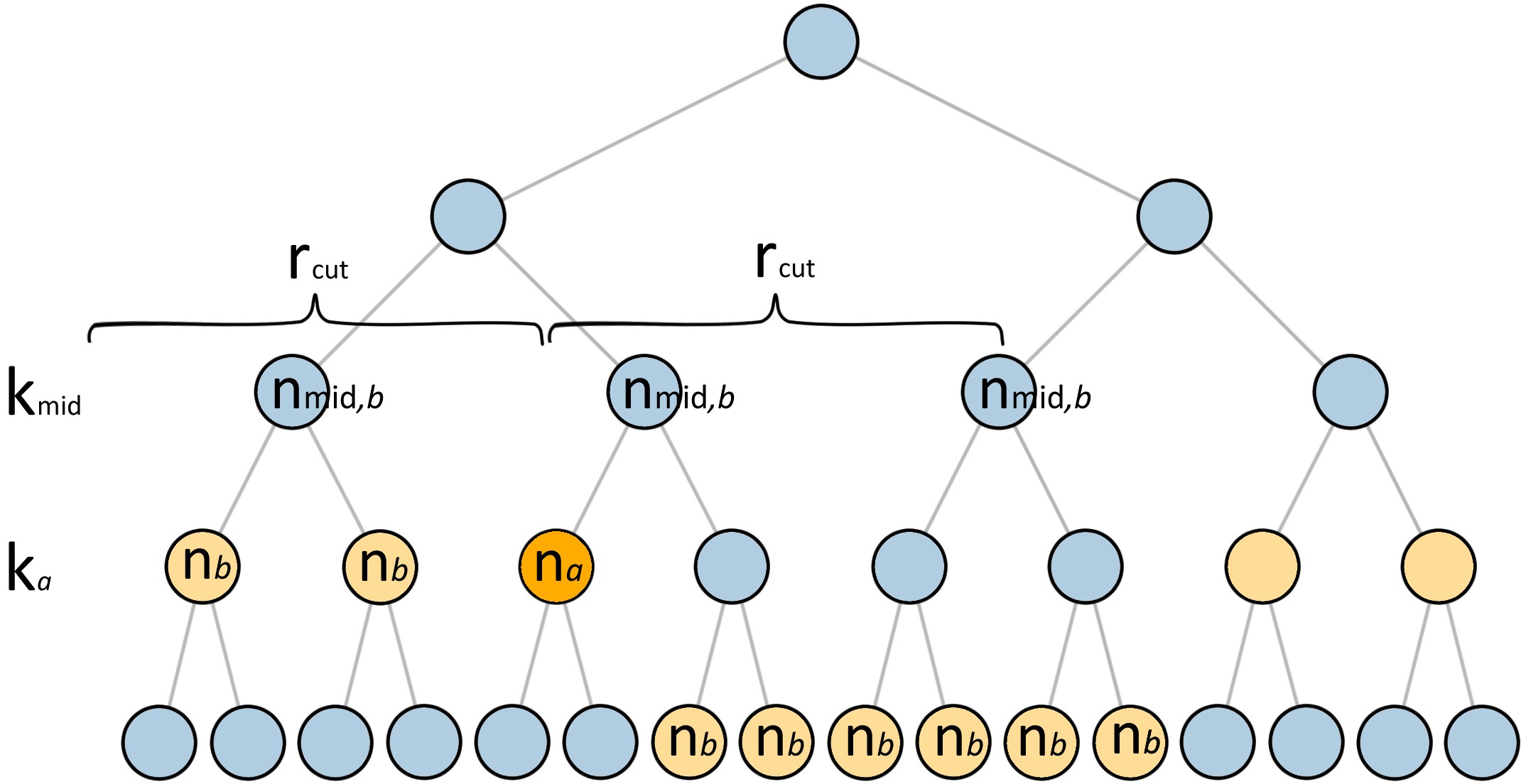}
    \caption{Illustration of the neighbour-find algorithm. Nodes in orange are example set of pseudo-particles. Consider an arbitrary target node \(n_a\), coloured in darker orange, located on level \(k_a\). Consider an arbitrary level above all pseudo-particles, labelled \(k_\mathrm{mid}\). For nodes on \(k_\mathrm{mid}\) that fall within the threshold radius \(r_\mathrm{cut}\) around \(n_a\), we label them \(n_{\mathrm{mid},b}\); their descendant pseudo-particles \(n_b\) are considered potential neighbours.  }
    \label{fig:neighfind}
\end{figure}

A fast neighbour-finding scheme is vital to the smoothing length solver. Typically, in astrophysical SPH codes, the gravity tree is traversed again for extracting neighbours, yet the fact that our pseudo-particles themselves are tree nodes precludes them from following this approach. Fortunately, the labelling system of our RCB tree allows node relations to be recovered. We make use of this property to locate their `distant relatives'. Fig. \ref{fig:neighfind} illustrates this algorithm. 

Consider a level somewhere midway on the tree, labelled \(k_\mathrm{mid}\). Level \(k_\mathrm{mid}\) must be above all pseudo-particles and, ideally, \textit{just} above the highest one. Next, consider a target pseudo-particle \(n_a\). We first evaluate the distance between \(n_a\) and all nodes on \(k_\mathrm{mid}\), labelled \(n_\mathrm{mid}\), and flag the ones that fall within a certain threshold radius \(r_\mathrm{cut}\). We then loop through the rest of the pseudo-particles \(n_b\). For each \(n_b\), we locate its ancestor on \(k_\mathrm{mid}\) using its index, which we label \(n_{\mathrm{mid},b}\). If \(n_{\mathrm{mid},b}\) was flagged in the previous procedure, we immediately add \(n_b\) to the trial neighbour list. 

Indeed, the key for this algorithm to operate accurately and efficiently lies in setting the right threshold \(r_\mathrm{cut}\). If \(n_a\) is low on the tree, the size of its ancestor on \(k_\mathrm{mid}\) (labelled \(s_{\mathrm{mid},a}\)) is usually sufficient to cover all of its neighbouring nodes. We make a more conservative estimate by taking the diagonal of the cell, giving \(r_\mathrm{cut} = \sqrt{3} s_{\mathrm{mid},a}\). Of course, \(n_a\) could be higher on the tree and thus closer to \(k_\mathrm{mid}\), in which case the above definition of \(r_\mathrm{cut}\) becomes insufficient. As such, we always perform a check afterwards by estimating the compact support radius of \(n_a\), which we denote \(r_{2h}\). For cubic spline kernels, this can be determined by simply taking the double of the smoothing length estimate in (\ref{eq:h_estimate}), hence
\begin{eqnarray} \label{eq:compact_support_radius}
    r_{2h} = 4 \ h_\mathrm{fact,node} \ s_a,
\end{eqnarray}
where \(s_a\) is the size of node \(n_a\). If the initially computed \(r_\mathrm{cut}\) is smaller than \(r_{2h}\), it likely indicates that \(n_a\) is high on tree and we require an alternative method for estimating the threshold. That said, this method follows the same principle. We consider the parent of \(n_a\) on one or two levels\footnote{The user can adjust this number if necessary, especially when the pseudo-particles are `steep' on the tree, i.e. the opening angle is large.} above \(k_a\), labelling it \(n_\mathrm{above}\). We again apply (\ref{eq:compact_support_radius}) but replacing the node size \(s_a\) by that of its parent, \(s_\mathrm{above}\). This conservative approximation takes into account the jumps in tree levels amongst neighbours and we set this to be the threshold \(r_\mathrm{cut}\), replacing the previous estimation.

The neighbour list is cached for fast retrieval. However, in situations where the total number of pseudo-particles is small, the overhead becomes prominent. We therefore activate this neighbour-search only if more than \(10^4\) nodes (excluding the individual SPH particles but counting the leaves) are extracted from the tree; otherwise, we resolve to a brute-force approach when solving for the smoothing lengths.

\section{Benchmarking} \label{benchmark}

\subsection{The {\scshape Starbench} tests} \label{starbench}

We now present the results from testing this RHD scheme against the well-established analytical solutions comprised in {\scshape Starbench}, a set of benchmarks compiled by \cite{bisbas11} for calibrating radiative transfer algorithms. This test problem examines the phase of HII region evolution during which the expansion is purely driven by the pressure imbalance between the hot ionized gas and the ambient uniform density medium. In this proceeding, we focus on the early-time behaviour.

Suppose a radiative source with ionizing photon flux \(Q\) is `switched on' in a uniform medium. The ionization front radius expansion can be derived by equating the ram pressure of the neutral gas to the pressure within the shell of shocked gas. This is known as the Spitzer solution\cite{spitzer78},
\begin{eqnarray} \label{eq:spitzer}
R_\mathrm{Sp}(t) = R_\mathrm{St} \left( 1+ \frac{7}{4} \frac{c_{i}t}{R_\mathrm{St}} \right) ^{4/7},
\end{eqnarray}
where \( c_{i} \) is the sound speed in the ionized medium. The Str\"{o}mgren radius\cite{stromgren39} \( R_\mathrm{St} \) in (\ref{eq:spitzer}) is given by 
\begin{eqnarray} \label{eq:stromgren_radius}
R_\mathrm{St} = \left( \frac{3Q}{4 \pi \alpha(H^0,T) n_H^2} \right) ^{1/3},
\end{eqnarray}
where \(\alpha(H^0,T)\) is the recombination coefficient of hydrogen as a function of temperature \(T\), and \(n_H\) is the total number density of hydrogen. This radius defines the boundary of the ionized region immediately upon reaching ionization equilibrium. Equation (\ref{eq:spitzer}) was later modified by \cite{hosogawainutsuka06} to take into account of the inertia of the expanding gas shell, giving the Hosokawa-Inutsuka solution,
\begin{eqnarray} \label{eq:hosokawa_inutsuka}
R_\mathrm{HI}(t) = R_\mathrm{St} \left( 1+ \frac{7}{4} \sqrt{\frac{4}{3}} \frac{c_{i}t}{R_\mathrm{St}} \right) ^{4/7}.
\end{eqnarray}

Our simulations closely follow the {\scshape Starbench} setup. We do not include gravity, turbulence or any other external forces in order to isolate the effect of thermal pressure inside the HII region. We set \( Q = 10^{49} \ \mathrm{s^{-1}} \). The photons are monochromatic with energy \( h \nu = 13.6 \ \mathrm{eV}\). The ambient medium has temperature \(T_0 = 10^2 \ \mathrm{K}\) and density \(\rho_0 = 5.21 \times 10^{-21} \ \mathrm{ g \ cm^{-3}}\), with which \(n_{H}\) can be determined assuming a pure hydrogen composition. For consistency with {\scshape Starbench}, we fix the temperature of ionized particles to \(T_i = 10^4 \ \mathrm{K}\). Re-emission of ionizing photons are neglected in this test. 

\begin{figure}
    \centering
    \includegraphics[width=3.4in]{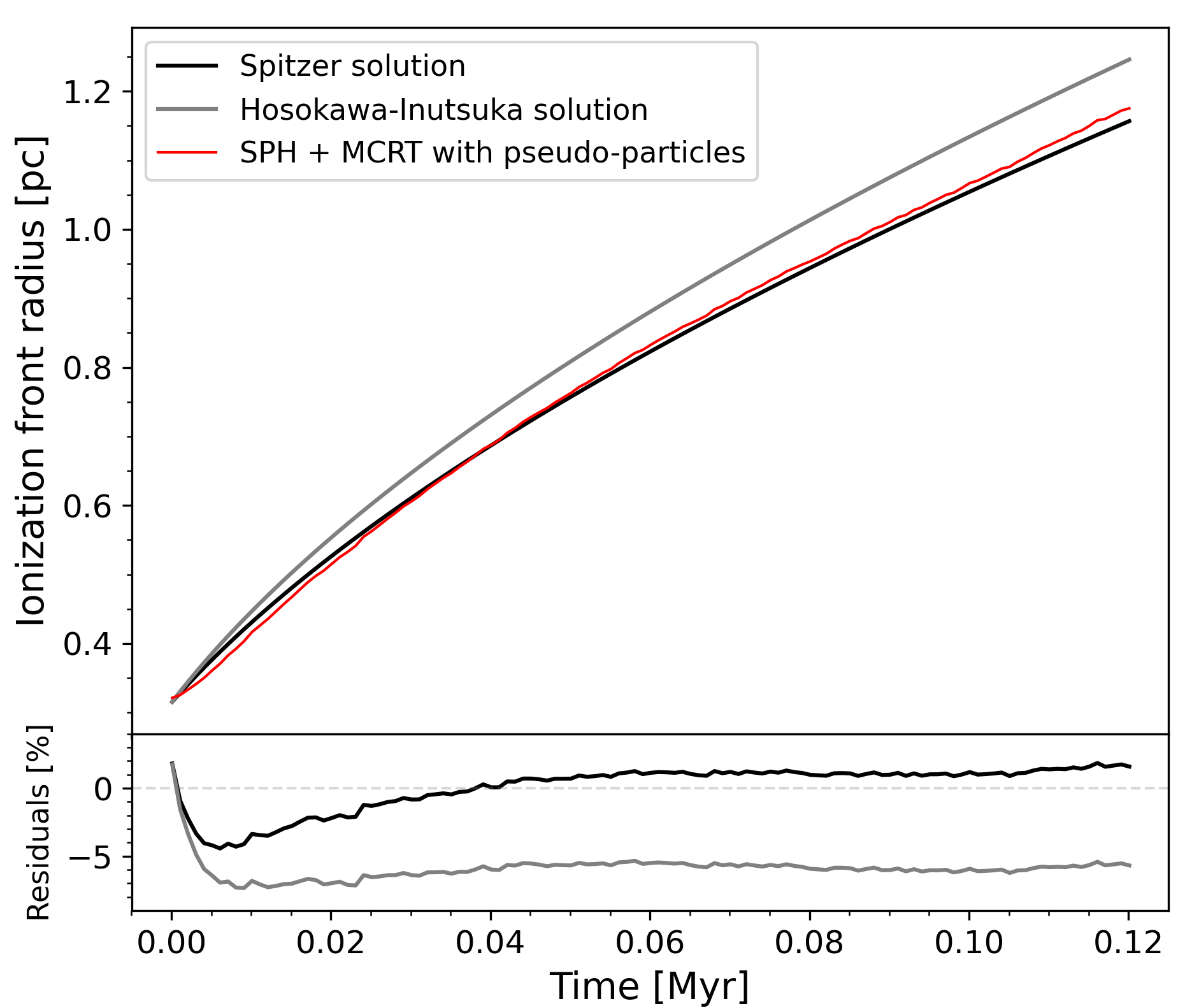}
    \caption{Top panel shows the evolution of the HII region ionization front radius from our {\scshape Starbench} early-phase benchmark test. Analytical solutions (\ref{eq:spitzer}) and (\ref{eq:hosokawa_inutsuka}) are plotted in black and grey respectively. Red curve shows the results from using our RHD scheme ran with pseudo-particles. Percentage errors relative to the analytical solutions, calculated with \( \left( R_\mathrm{sim} - R_\mathrm{sol} \right) / R_\mathrm{sol} \times 100  \), are presented in the bottom panel. Time in \(\mathrm{Myr}\) \((3.156 \times 10^{13} \ \mathrm{s})\).}
    \label{fig:starbench}
\end{figure}

Fig. \ref{fig:starbench} compares the results produced with our RHD scheme using pseudo-particles against (\ref{eq:spitzer}) and (\ref{eq:hosokawa_inutsuka}). We set \(K = 100\) in (\ref{eq:resol_param}) such that the ionized region is always resolved by individual particles (c.f. Fig. \ref{fig:node_illus} bottom-right panel). Our result agrees well with the analytical solutions, especially (\ref{eq:spitzer}). It also matches the findings reported in \cite{petkova21} and \cite{bisbas15}. Our results demonstrate that using pseudo-particles immediately beyond the ionization front does not hamper the simulation accuracy.

\subsection{Speed-up} \label{speedup}

\begin{figure}
    \centering
    \includegraphics[width=3.0in]{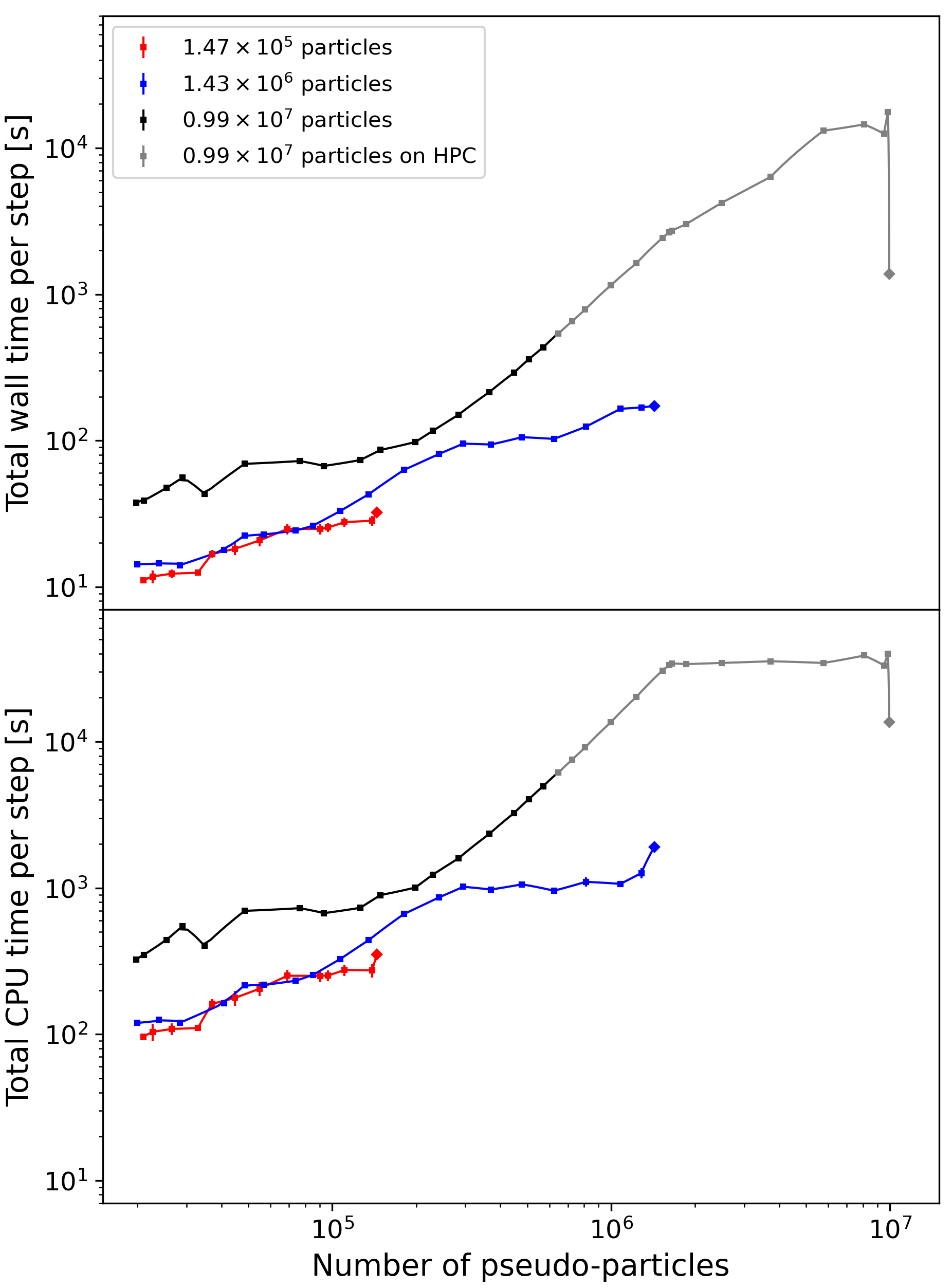}
    \caption{Wall-time (top panel) and CPU-time (bottom panel) per step against number of pseudo-particles, plotted for simulations with \(\sim 10^5\) particles (red), \(\sim 10^6\) particles (blue) and \(\sim 10^7\) particles (black and grey for those run on HPC). Diamonds indicate the runs without optimization. Error bars indicate the standard deviation in gradient of the elapsed time across 6 steps. }
    \label{fig:runtime_total}
\end{figure}

To illustrate the performance of this optimization algorithm, this section presents the runtimes of the {\scshape Starbench} test. While keeping the HII region fully resolved, we vary the number of pseudo-particles at the coupling interface to gauge the amount of speed-up that can be achieved. The runtimes are measured on a 6-core 2200 MHz processor machine with 12 threads. For tests with large (pseudo-)particles numbers, the simulations are run on a 16-core HPC cluster. We afterwards scale their runtimes to allow for comparisons with the 6-core machine. Fig. \ref{fig:runtime_total} shows the results after applying this correction. 

\begin{figure}
    \centering
    \includegraphics[width=3.0in]{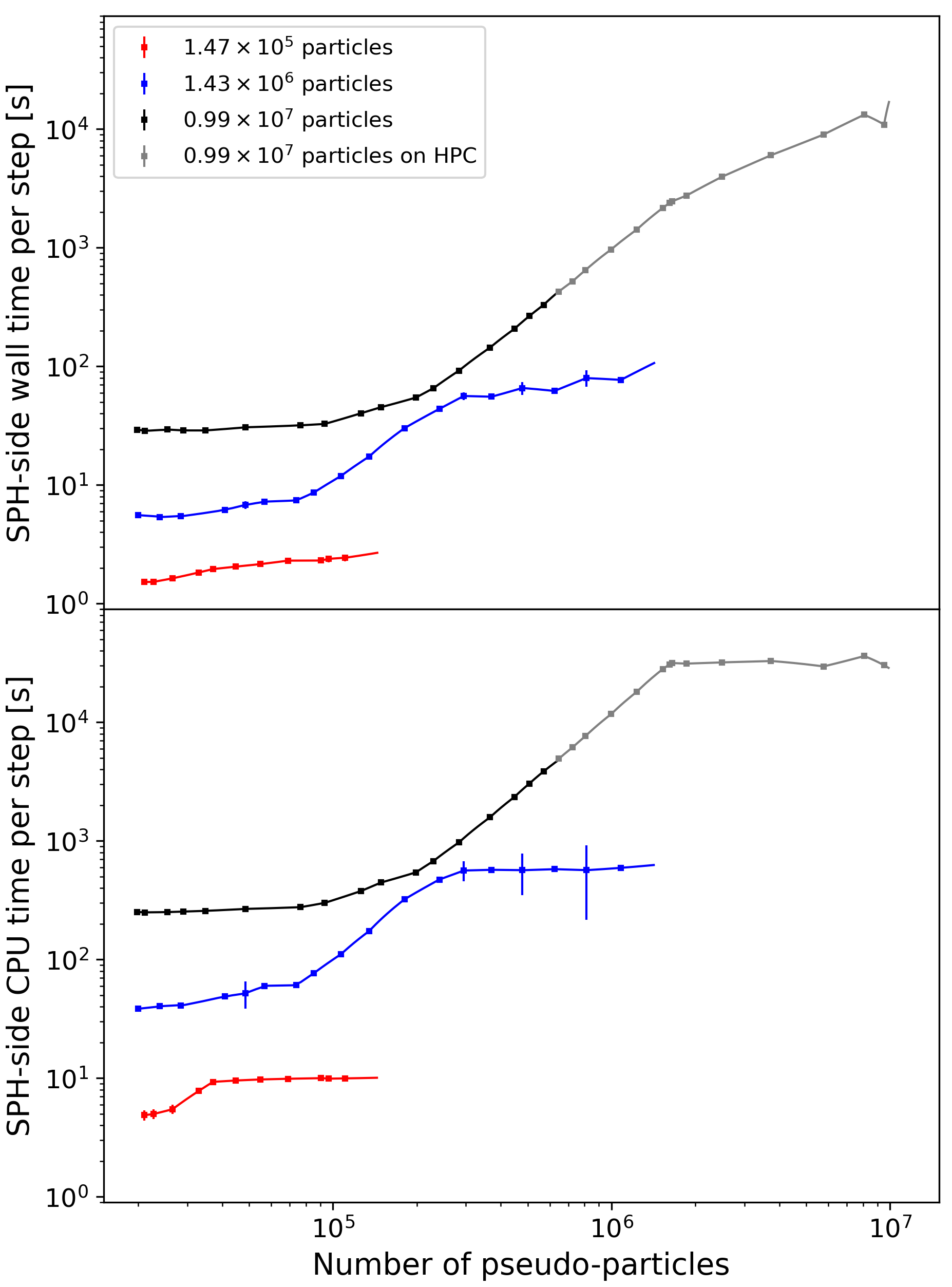}
    \caption{Same as Fig. \ref{fig:runtime_total} but runtimes include only the contribution from the SPH side of this RHD scheme, excluding the time for the MCRT simulation, grid-construction and density-mapping. The curves exclude the data points for when no optimizations are applied. }
    \label{fig:runtime_sph_side_only}
\end{figure}

We performed three sets of {\scshape Starbench} simulations, each with approximately \(10^5\), \(10^6\) and \(10^7\) particles\footnote{The particle mass and the density of the ambient medium are the same for all cases, hence box sizes increase with number of particles.}. The number of pseudo-particles is varied for each test by tweaking \(\theta\), \(r_\mathrm{part}\) and \(r_\mathrm{leaf}\). In this simulation, at least \(10^4\) pseudo-particles are required to keep the ionized region resolved. We run each test for 6 timesteps and obtain the average runtime per step by fitting a linear regression to the recorded instants. Fluctuations along the curves in Fig. \ref{fig:runtime_total} are caused by minor inconsistencies in the number of cells merged. 

We determine the amount of speed-up by comparing the run with no optimizations applied (last data point on each curve) to the run that uses minimum number of pseudo-particles (first data point). For the case with \(10^5\) particles, the CPU-time is reduced by around a factor of 3. This fractional difference rises to 20 for \(10^6\) particles, and 30 for \(10^7\) particles. It is evidenced that the amount of speed-up scales positively with the reduction in number of particles at the coupling interface. 

One issue to note from Fig. \ref{fig:runtime_total} is the plateauing of CPU-time. We found that this scaling behaviour originates from the optimization algorithm (see Fig. \ref{fig:runtime_sph_side_only} which isolates the runtime of the SPH code) - more specifically, the choice of \(r_\mathrm{leaf}\) and \(r_\mathrm{part}\). Here in our tests, in order to reach the large pseudo-particle number regime, \(r_\mathrm{leaf}\) must cover the whole space and only \(r_\mathrm{part}\) is varied. Yet, recall from Section \ref{h_solver} that our code only computes smoothing lengths for tree nodes. As \(r_\mathrm{leaf}\) was constant, the load handled by the smoothing length solver in fact remained the same. This leads to the flattening of CPU-time and at the same time shows the dominance of smoothing length computation when the number of pseudo-particles is immense. Their runtimes exceed the case with no optimizations applied. This overhead is worsened by a serial procedure which is responsible for removing the temporarily inserted leaves around the ionizing source. Its scaling with \(r_\mathrm{part}\) invoked the rise in wall-time. 

We stress, however, that such issues arise only when the number of pseudo-particles approaches the total number of SPH particles. In practice, this number would only be minimized. Further, the peculiar treatments associated with \(r_\mathrm{part}\) and \(r_\mathrm{leaf}\) would not have been needed if the tree is built down to particle level. Hence, these overheads are inconsequential. The minimal runtime reached is our sole concern.

\section{Applications to realistic cloud simulations} \label{applications}

\begin{figure}
    \centering
    \includegraphics[width=3.45in]{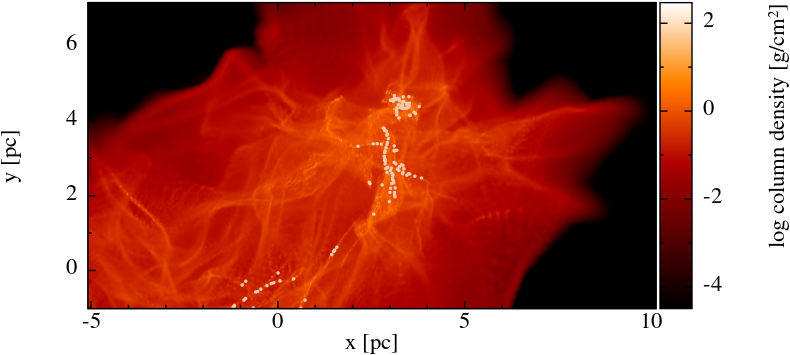}
    \caption{Column density of a turbulent \(10^5 \ \mathrm{M_{\odot}}\) Giant Molecular Cloud with warm envelope evolved for \(0.38 \ \mathrm{Myr}\). White dots indicate locations of sink particles (stars). The ionizing source is located at \((2.8 \ \mathrm{pc}, \ 3.0 \ \mathrm{pc})\) on this projection. Image produced with {\scshape Splash} \cite{splash07}. }
   \label{fig:coldens_cloud_beforeinject}
\end{figure}

Fig. \ref{fig:coldens_cloud_beforeinject} shows a column density plot from a simulation of a \(10^5 \ \mathrm{M_{\odot}}\) (\(1.989 \times 10^{35} \ \mathrm{kg}\)) turbulent Giant Molecular Cloud with \(10^6\) gas particles at around a temperature of 10 K, modelled with an adiabatic equation of state. The mean density is approximately \( 10^{-21} \ \mathrm{g \ cm^{-3}}\). We enveloped the cloud with warm gas at 1000 K. The cloud was evolved for 0.38 Myr and formed approximately 170 sink particles\cite{bate95} that represent individual stars. To demonstrate the effect of a single HII region, we selected the most massive sink to be the ionizing source. We assumed an ionizing flux of \(Q = 10^{51} \ \mathrm{s}^{-1}\) which corresponds to the output from a massive stellar cluster. Our RHD scheme is then activated to heat the ionized particles.

\begin{figure}
    \centering
    \includegraphics[width=3.45in]{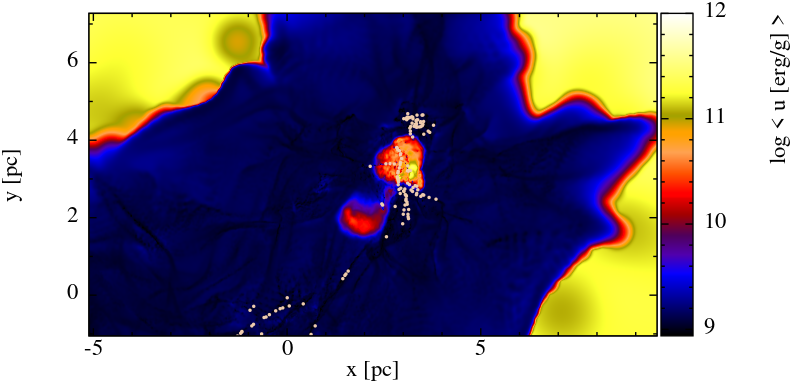}
    \caption{Internal energy of the Giant Molecular Cloud after incorporating photoionization. White dots indicate locations of sink particles. Energy in units of \(\mathrm{erg/g}\) (\(10^{-4}\ \mathrm{J/kg}\)). Image produced with {\scshape Splash} \cite{splash07}. }
    \label{fig:IE_cloud_afterinject}
\end{figure}

As the cloud evolved, the HII region rapidly reached its equilibrium temperature. Fig. \ref{fig:IE_cloud_afterinject} shows the gas internal energy \(4.70 \times 10^{-3} \) Myr
after switching on the ionizing source. A hot bubble is seen emerging from the densest filaments of the cloud where the source is located. The maximum internal energy of this bubble is approximately three orders of magnitude higher than that of the cloud, in agreement with our initial conditions and the typical temperature of HII regions. We also see a glob of warm gas at some distance away from the HII region, which appears to have escaped the confinement and is reshaping the surrounding medium.

\section{Conclusion} \label{conclusion}

We presented an RHD scheme that couples SPH to grid-based MCRT to incorporate photoionization in astrophysical simulations. We applied the Exact density-mapping, an analytical solution to the volume integral of SPH interpolation formula, to transfer fluid densities between particles and Voronoi grid cells. To optimize the coupling, we developed methods to turn gravity tree nodes into pseudo-particles, reducing resolution in non-ionized regions. In our runtime tests, we achieved a speed-up of 20 times for \(10^6\) particles and 30 times for \(10^7\) particles. The maximal speed-up depends on the size of the HII region relative to the whole simulation box. 

For photoionization, it indeed can be redundant to consider the pseudo-particles far beyond the ionization front as ionic fractions rapidly drop to zero. This tree-based method is arguably even more suitable for modelling physical mechanisms whose influence scales inversely with distance, such as radiation pressure, which is also important in driving molecular cloud evolution. This could be included with the help of MCRT simulations to compute the momentum transfer from photons onto SPH particles, and place pseudo-particles based on the amount of force received. Our optimization algorithm can be implemented in any SPH/\textit{N}-body codes as long as a geometrical tree is in place. Together with the density-mapping, the techniques presented here are particularly useful for coupling SPH to other numerical methods.

\section*{Acknowledgment}
We thank James Wurster for all the very helpful advice on algorithms and comments that improved this paper. This work was supported by the STFC training grant ST/W507817/1. Part of the simulations were performed using the HPC Kennedy, operated by University of St Andrews.



%

\bibliographystyle{IEEEtran}
\bibliography{ref_list.bib}

\end{document}